\documentclass[9pt]{extarticle}

\usepackage{spconf,amsmath,graphicx}
\usepackage{epstopdf}
\usepackage{setspace}

\usepackage{graphicx}
\usepackage{color}
\usepackage{placeins}
\usepackage{float}
\usepackage{tabularx,colortbl}
\graphicspath{{figures/Appendix/}}
\usepackage{amssymb} % \in
\usepackage{amsmath} % For align command
\usepackage[breaklinks=true]{hyperref}
\usepackage{setspace}
\usepackage{cite}
\usepackage{subfigure}
\DeclareGraphicsExtensions{.pdf,.png,.jpg}

\title{Approximate Computation of DFT without Performing Any Multiplications: Applications to Radar Signal Processing}
\name{Alican Bozkurt, Musa Tun\c{c} Arslan, Rasim Akin Sevimli, Cem Emre Akbas, A. Enis \c{C}etin}
\address{Dept. of Electrical and Electronic Engineering, Bilkent University,  Ankara, Turkey\\
alican@ee.bilkent.edu.tr, mtarslan@ee.bilkent.edu.tr, sevimli@ee.bilkent.edu.tr, akbas@ee.bilkent.edu.tr, cetin@bilkent.edu.tr\\}

\begin{document}
%\ninept
%
\maketitle
\begin{abstract}

In many practical problems it is not necessary to compute the DFT in a perfect manner including some radar problems. In this article a new multiplication free algorithm for approximate computation of the DFT is introduced. All multiplications $(a\times b)$ in DFT are replaced by an operator which computes $sign(a\times b)(|a|+|b|)$.  The new transform is especially useful when the signal processing algorithm requires correlations. Ambiguity function in radar signal processing requires high number of multiplications to compute the correlations. This new additive operator is used to decrease the number of multiplications. Simulation examples involving passive radars are presented.
\end{abstract}
%
%\keywords{one, two, three, four}
\section{Introduction}
\label{sec:Introduction}

In \cite{tuna2009image, alex2013multip}, we introduced a new vector "product" and the codifference operator as a computationally efficient alternative to the commonly used inner-product and covariance operators. The new vector product and the codifference operator are based on replacing multiplication by an additive operation:

\begin{equation}
\label{eq:1}
a\otimes b = sign(a\times b)(|a|+|b|),
\end{equation}
where,
\begin{equation}
\label{eq:2}
sign(a\times b) = \begin{cases}
	 1, & \text{if $a > 0, b > 0$ or $a < 0, b < 0$},\\
	-1, & \text{if $a < 0, b > 0$ or $a > 0, b < 0$},\\
     0, & \text{otherwise}.
\end{cases}
\end{equation}

The vector product of two vectors $\textbf{x}$, $\textbf{y} \in R^N$ is defined as follows,
\begin{equation}
\label{eq:3}
<\textbf{x}\odot \textbf{y}> = \sum_{i=1}^{N}x_i\otimes y_i,
\end{equation}
where $x_i$ and $y_i$ are the $i^{th}$ components of vectors $\textbf{x}$ and $\textbf{y}$, respectively. When the means of vectors \textbf{x} and \textbf{y} are equal to zero, $<\textbf{x}\odot \textbf{y}>$ is called the codifference of vectors \textbf{x} and \textbf{y} \cite{alex2013multip}.

We define the multiplication of a vector \textbf{x} by a number $a\in \mathbb{R}$ based on Equation \ref{eq:1} and \ref{eq:3} as follows:
\begin{equation}
\label{eq:4}
 a \odot \textbf{x}=\begin{bmatrix} a \odot {x}(1) \quad  a \odot {x}(2) \hdots  a \odot {x}(N)\end{bmatrix}^T,
\end{equation}
where $a$ is an arbitrary real number. Note that the vector product of a vector $\textbf{x}$ with itself reduces to a scaled $l_1$ norm of \textbf{x} as follows:
\begin{equation}
<\textbf{x} \odot \textbf{x}>=\sum\limits_{i=1}^N {x}(i) \odot {x}(i)=2\sum\limits_{i=1}^N |{x}(i)|=2||{x}||_{1},
\label{eq:5}
\end{equation}

In this article the above vector product concept is used to approximately compute the DFT.
In Section 2, the new nonlinear transform approximating DFT is described. In Section 3, the FFT version of the nonlinear DFT is described. The nonlinear FFT is designed based on the scattering approach introduced in \cite{lecun2010convolutional, mallat2012comm, huang2006large}. In Section 4, application of the nonlinear DFT to radar signal processing is presented. Experimental results are in Section 5.

\section{Nonlinear Transform Approximating DFT}
\label{sec:2}
In \cite{tuna2009image, alex2013multip}, real signals and images are used in vector product and codifference operations. We need to extend the additive operator defined in Equation (1) to complex numbers. Let $a$ and $b$ be two arbitrary complex numbers, $a\otimes b$ is defined as follows:
\begin{equation}
\label{eq:6}
\begin{split}
a\otimes b \triangleq &(a_r+ja_i)\otimes (b_r + jb_i) \\ = &a_r \otimes b_r - a_i\otimes b_i + j(a_i\otimes b_r + b_i\otimes a_r),
\end{split}
\end{equation}
where $a_r$, $b_r$ and $a_i$, $b_i$ are the real and the imaginary parts of $a$ and $b$, respectively.

It is possible to replace the matrix-vector product of DFT with Equation ($\ref{eq:6}$) to obtain a multiplication-free transform. This new transform is called Nonlinear DFT (NDFT).
Based on Equation (6), we define the NDFT of $x[n]$, $n=0,1,...,N-1$ as follows:
\begin{equation}
\label{eq:8}
\begin{pmatrix}
X[0] \\ X[1]\\ \vdots \\ X[N-1]
\end{pmatrix}
=
\begin{pmatrix}
1 &1 &\cdots &1 \\
1 &W &\cdots &W^{N-1} \\
1 &W^2 &\cdots &W^{N-2} \\
\vdots &\vdots &\ddots &W^{N-3}\\
1 &W^{N-1} &\cdots &W
\end{pmatrix}
\odot
\begin{pmatrix}
x[0] \\
x[1] \\
\vdots \\
x[N-1]\\
\end{pmatrix}
\end{equation}
where $W=e^{-j2\pi /N}$, and $X[0], ... , X[N-1]$ are the NDFT coefficients.

Let $x[n] = e^{j2\pi k_on/N}$, the NDFT has a peak because of the following inequality:
\begin{equation}
\begin{split}
\label{eq:9}
|<x[n] \odot &[1 ... e^{-j2\pi k_0n/N}... e^{-j2\pi k_0(N-1)/N}]>|\quad \geq \quad\\ &|<x[n] \odot [1 ... e^{-j2\pi kn/N} ... e^{-j2\pi k(N-1)/N}]>|,
\end{split}
\end{equation}
for $k \neq k_0$.

An example NDFT computation for $x[n] = e^{j2\pi 7n/N}$, $n = 0,...,N-1$, is plotted in Fig. 1 for NDFT size of $N=64$. The peak is clearly visible at $k=7$, but it is slightly scattered across the $k$ values due to the nonlinear nature of $\odot$.
\begin{figure}[h!]
\label{fig:1}
  \centering
    \includegraphics[width=0.5\textwidth]{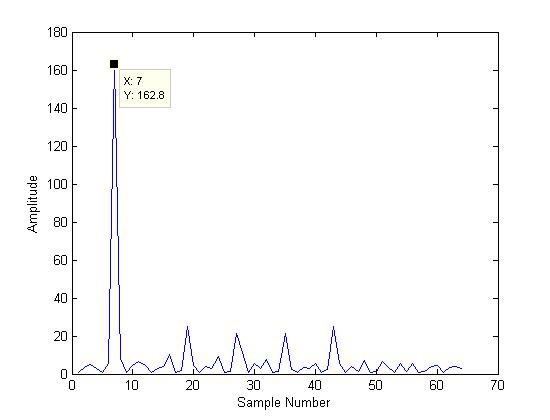}
    \caption{NDFT of $x[n] = e^{j2\pi 7n/N}$, $n = 0,...,N-1$, $N=64$.}
\end{figure}

The N-point NDFT has computational cost of $N^2$ complex $\odot$ operations and additions. Each $\odot$ requires $4$ sign computations $8$ real absolute value computations and $6$ additions.

The NDFT transform defined in Equation (7) is a multiplication free transform, but it is relatively slow. Another family of nonlinear transforms can be defined by replacing the complex multiplications with additive operator $\odot$ in decimation-in-time and decimation-in-frequency FFT algorithms. This approach is similar to the scattering approach used in \cite{sifre2012combined, hadsell2006dimensionality}. Application of the additive operator $\odot$ to FFT algorithms is described in Section 3.

\section{Nonlinear Transform Approximating FFT}
In this section, Nonlinear FFT (NFFT) is defined as in NDFT. The nonlinear FFT is designed similar to the scattering wavelet transform by replacing a linear operator (multiplication) with a nonlinear operator \cite{anden2011multiscale, bruna2012invariant, boureau2010learning}.

By far Cooley-Tukey approach is the most commonly used FFT algorithm to compute DFT. Radix-2 decimation-in-time algorithm, uses a divide-and-conquer type of approach to rearrange an N-point DFT into two parts as follows:
\begin{equation}
\begin{split}
X_{DFT}[k] &= \sum_{n=0}^{N/2-1}x[2n]W_N^{2kn} + W_N^k\sum_{n=0}^{N/2-1}x[2n+1]W_N^{2kn}
\end{split}
\end{equation}
for $k=0,1,2,...,N-1$ and $W_N^{2kn} =e^{-j2\pi 2nk/N}$.
The first sum is the $N/2$-point DFT of even indexed $x[n]$ and the second sum is the $N/2$-point DFT of odd indexed $x[n]$ coefficients.
Both of these $N/2$-point DFTs can be rearranged again to get four $N/4$-point DFTs and this process can be repeated until $2$-point DFTs remain. Therefore, Fourier transform is calculated using multiple $2$-point DFT butterflies. Decimation-in-frequency based NFFT can be defined in a similar manner.

\begin{figure}[h!]
\label{fig:2}
  \centering
    \includegraphics[width=0.5\textwidth]{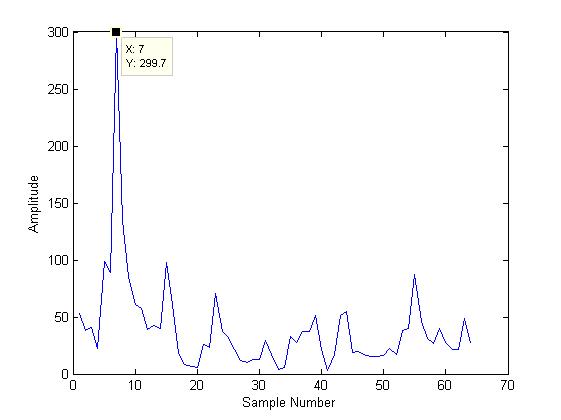}
    \caption{NFFT $\hat{X}[k]$, $k=0,1,...,N-1$ of $x[n] = e^{j2\pi 7n/N}$, $n = 0,...,N-1$, $N=64$.}
\end{figure}

For each DFT step in FFT, the additive operator defined in Equation (\ref{eq:6}) can be used. Thus, a multiplication-free Nonlinear FFT (NFFT) is obtained based on the following equation:
\begin{equation}
\begin{split}
\hat{X}[k] = \sum_{n=0}^{N/2-1}x[2n]\odot W_N^{2kn} + W_N^k\sum_{n=0}^{N/2-1}x[2n+1]\odot W_N^{2kn}
\end{split}
\end{equation}
for $k=0,1,2,\ldots,N-1$ and $W_N^{2kn} =e^{-j2\pi 2nk/N}$.
An example NFFT for $x[n] = e^{j2\pi 7n/N}$, $n = 0,\ldots,N-1$, NFFT size $N=64$ is in Fig.~\ref{fig:2}.

The N-point NFFT has computational cost of $NlogN$ complex $\odot$ operations. Each $\odot$ requires, $4$ sign computations $8$ real absolute value computations and $6$ additions.

Since the operator $\odot$ is a nonlinear operator, NFFT values $X_F[k]$, $k=0,1,...,N-1$ are not equal to NDFT values obtained in Equation (7).

\section{Application to Radar Signal Processing}
FFT can be used to overcome various problems in radars \cite{gedney1990use}. In this section, we apply the NDFT and NFFT presented in Section 2 and 3, respectively, to radar signal processing.

In practice, transmitted signal echoes off from obstacles in the environment. Moving obstacles are considered as targets to be detected, stationary obstacles are considered as clutters. Target echoes are time delayed and Doppler-shifted. Clutter echoes are only time delayed.
The ambiguity function used in detection of targets and clutters in radar signal processing is defined as follows:
\begin{equation}
\label{eq:10}
A[l,p] = \sum_{i=0}^{N-1}{s_{surv}[i]s_{ref}^*[i-l]e^{-j2\pi ip/N}}
\end{equation}
where $A[l,p]$ is the amplitude range-doppler surface, $l$ is the range bin of interest, $p$ is the Doppler bin of interest, $s_{surv}[i]$ is the surveillance radar input, $s_{ref}[i]$ is the reference radar input \cite{colone2009multistage}.

Implementation of this equation is simply calculating the N-point Discrete Fourier Transform of the $s_{surv}[i]\times s_{ref}^*[i-l]$, for $i=0,1,...,N-1$. Since the  baseband signals are used in the radar, $s_{surv}[i]$ and $s_{ref}[i]$ are complex valued signals. Therefore, by using the extended additive operator in Equation (6) it is possible to obtain new ambiguity functions.

We define three new ambiguity functions based on the nonlinear operator $\odot$ as follows:
\begin{subequations}
\begin{equation}
\label{eq:12a}
\bar{A}[l,p] = \sum_{i=0}^{N-1}s_{surv}[i]\odot s_{ref}[i-l]\odot e^{-j2\pi ip/N}
\end{equation}
\begin{equation}
\label{eq:12b}
\hat{A}[l,p] = \sum_{i=0}^{N-1}(s_{surv}[i]\odot s_{ref}[i-l])\times e^{-j2\pi ip/N}
\end{equation}
and,
\begin{equation}
\label{eq:12c}
\widetilde{A}[l,p] = \sum_{i=0}^{N-1}(s_{surv}[i]\times s_{ref}[i-l])\odot e^{-j2\pi ip/N}
\end{equation}
\end{subequations}

In the next section we present simulation examples using (12a) and (12b). Equation (12c) showed relatively inferior results compared to Equation (12a) and (12b). As a result, we do not present any simulation studies using $\widetilde{A}[l,p]$.

\section{Simulation Results}

In recent years, with the development of DSP algorithms, passive bistatic radars (PBR) for surveillance purposes renewed interest. PBR exploits an existing commercial transmitter as an illuminator of opportunity. Most common signals in use today for PBRs are commercial FM radio transmitters \cite{baker2005passive}. High transmit power of FM transmitters especially make them useful for detection of long range targets \cite{howland2005fm}.

Simulations are carried out in Matlab. First a stereo FM signal is generated as transmit signal, then, an environment is set up with various numbers of targets and clutters. The ambiguity function of the environment is computed by using the new ambiguity functions defined in Equation (12a) and (12b).

\subsection{Stereo FM Signal}

A baseband stereo FM signal has 200 kHz bandwidth with 100 kHz message bandwidth. Stereo FM message signal has the following form:
\begin{equation}
\label{eq:11}
\begin{split}
m(t) = 0.9(x_1 + x_2) &+ 0.5(x_1-x_2)cos(2\pi 2f_pt) \\ &+0.25cos(2\pi 3f_pt) + 0.1cos(2\pi f_pt)
\end{split}
\end{equation}
where $x_1$ and $x_2$ are the left and right channel information, respectively. They are randomly generated numbers to represent the random-like behaviour of a typical FM broadcast. $f_p$ is the pilot tune frequency set at 19 kHz. The message signal $m(t)$ is modulated as follows:
\begin{equation}
\label{eq:12}
s(t) = cos(2\pi k_fm(t))+jsin(2\pi k_fm(t))
\end{equation}
where $k_f$ is the modulation index. Naturally, baseband FM signals are complex valued. They are sampled at sampling frequency $f_s = 200 kHz$.

In radar signal processing, moving objects are called as targets, which create time delay and Doppler shift on the transmit signal. Stationary objects are called as clutters, which only create time delay on transmit signal. A typical far field echo signal is as follows:
\begin{equation}
\label{eq:13}
s_{target}(t) = Ks(t-t_0)\times e^{j2\pi f_dt}
\end{equation}
where $K$ is the bistatic radar equation attenuation, $s(t-t_0)$ is the time delayed FM signal, $f_d$ is the Doppler shift of the target. Clutter echo signals are the same as Eq. (\ref{eq:13}), but with $f_d = 0$, because clutters are stationary obstacles.

Lastly, all target and clutter echo signals are summed to generate the surveillance signal $s_{surv}(t)$. For an environment with $n_t$ targets and $n_c$ clutters surveillance signal is generated as follows:
\begin{equation}
s_{surv}(t) = \sum_{m=1}^{n_t+n_c}a_ms_{ref}(t-\tau_m)e^{j2\pi f_{dm}t}
\end{equation}
where $a_m$, $\tau_m$ and $f_{dm}$  are the complex amplitude, the delay, and Doppler frequency of the $m$-th obstacle. For clutters, $f_{dm} = 0$.

\subsection{Detection of Targets and Clutters}
In this section, target and clutter objects are detected using the ambiguity functions defined in Equations (12a) and (12b). $s_{surv}(t)$ is generated using the Eq. (16). It is assumed that $s_{ref} = s(t)$. As a result, the ambiguity functions detect the bistatic ranges instead of the actual ranges of target and clutter objects. For the simulation purposes, additive white Gaussian noise and epsilon contaminated Gaussian noise is considered. The reference signal $s_{ref}$ is assumed to be noise free as in \cite{colone2009multistage}.

Since NFFT is a nonlinear operation input level is important for detection performance. The signal $s_{ssurv}$ is amplified $64$ times and NFFT is amplified $16$ times in Table 1.

In general FFT based ambiguity function provides better side-lobe performance as it can be seen from Tables 1 and 2 compared to the nonlinear FFT. However, FFT based ambiguity function fails to detect targets when the noise is heavy tailed (contaminated Gaussian). This is expected because $\odot$ operator induces the $l_1$ norm as described in Eq. (5). It is well-known that the $l_1$ norm based systems and algorithms are more robust to outliers compared to the Euclidean norm based correlation algorithms, \cite{tuna2009image, alex2013multip, kwak2008principal, ke2005robust}.
\section{Conclusion}
In this paper, a new computationally efficient approach to approximate DFT
based on a nonlinear additive operator is introduced. The main advantage of
this approach is that it enables a multiplication-free approximation of
DFT.  The nonlinear DFT can be computed using an FFT like algorithm.  The
cost of the N-point nonlinear FFT algorithm is, $4\times NlogN$ sign computations, $8\times NlogN$ absolute value computations and $6\times NlogN$ additions.

The nonlinear DFT is successfully used to compute the radar ambiguity
function and locate the peaks due to moving targets. It produces superior performance under contaminated Gaussian noise compared to ordinary ambiguity function.

In this problem it is
neither possible to use the wavelet transform nor the Hadamard transform
because the observed signal contains  Doppler terms due to moving objects.

\begin{table}[ht!]
\begin{center}
\caption{Simulation results for different environment and noise cases for nonlinear NFFT based ambiguity function. Eq. (12a). Rows 1,2,5,6,9 and 10 are AWGN and rows 3,4,7,8,11 and 12 correspond to additive contaminated Gaussian. Same applies to Table 2.}
\scalebox{0.8} {
\label{tab:1}
\begin{tabular}{|l|l|l|l|l|}
\hline
  \parbox[t]{1.34cm}{Environment} & \parbox[t]{1.34cm}{Performance}  & \parbox[t]{1.34cm}{Noise} & \parbox[t]{1.34cm}{Side-lobe \\ Floor (dB)} \\ \hline
  2 targets 1 clutter & detected & 3 dB& -3.86 \\ \hline
  2 targets 1 clutter & detected & 6 dB& -4.12 \\ \hline
  2 targets 1 clutter & detected & \parbox[t]{1.4cm}{\fontsize{5}{4}\selectfont eps. cont. \\ $\epsilon = 0.9$ \\ $\sigma_1=0.25$ \\$\sigma_2 = 10$} & -2.24 \\ \hline
  2 targets 1 clutter & detected & \parbox[t]{1.4cm}{\fontsize{5}{4}\selectfont eps. cont. \\ $\epsilon = 0.8$ \\ $\sigma_1=0.5$ \\$\sigma_2 = 20$} & -2.02 \\ \hline
  4 targets 2 clutters & 1 target masked & 3 dB& -3.34 \\ \hline
  4 targets 2 clutters & detected & 6 dB& -3.92 \\ \hline
  4 targets 2 clutters & \parbox[t]{1.34cm}{1 target\\1 clutter\\ masked} & \parbox[t]{1.4cm}{\fontsize{5}{4}\selectfont eps. cont. \\ $\epsilon = 0.9$ \\ $\sigma_1=0.25$ \\$\sigma_2 = 10$} & -2.99 \\ \hline
  4 targets 2 clutters & \parbox[t]{1.34cm}{2 clutters\\ masked} & \parbox[t]{1.4cm}{\fontsize{5}{4}\selectfont eps. cont. \\ $\epsilon = 0.8$ \\ $\sigma_1=0.5$ \\$\sigma_2 = 20$} & -2.57 \\ \hline
  1 target 3 clutters &	detected & 3 dB& -4.01 \\ \hline
  1 target 3 clutters &	detected & 6 dB& -3.87 \\ \hline
  1 target 3 clutters &	1 clutter masked & \parbox[t]{1.4cm}{\fontsize{5}{4}\selectfont eps. cont. \\ $\epsilon = 0.9$ \\ $\sigma_1=0.25$ \\$\sigma_2 = 10$} & -2.53 \\ \hline
  \hline
  1 target 3 clutters &	1 clutter masked & \parbox[t]{1.4cm}{\fontsize{5}{4}\selectfont eps. cont. \\ $\epsilon = 0.8$ \\ $\sigma_1=0.5$ \\$\sigma_2 = 20$} & -2.17 \\ \hline
  \hline
\end{tabular} }
\end{center}
\end{table}

\begin{table}[ht!]
\begin{center}
\caption{Simulation results for different environment and noise cases for FFT based ambiguity function. Eq. (11). }
\scalebox{0.8} {
\label{tab:2}
\begin{tabular}{|l|l|l|l|l|}
\hline
  \parbox[t]{1.34cm}{Environment} & \parbox[t]{1.34cm}{Performance}  & \parbox[t]{1.34cm}{Noise} & \parbox[t]{1.34cm}{Side-lobe \\ Floor (dB)} \\ \hline
  2 targets 1 clutter & detected & 3 dB& -5.98 \\ \hline
  2 targets 1 clutter & detected & 6 dB& -6.13 \\ \hline
  2 targets 1 clutter & no detection & \parbox[t]{1.4cm}{\fontsize{5}{4}\selectfont eps. cont. \\ $\epsilon = 0.9$ \\ $\sigma_1=0.25$ \\$\sigma_2 = 10$} & -0.57 \\ \hline
  2 targets 1 clutter & no detection & \parbox[t]{1.4cm}{\fontsize{5}{4}\selectfont eps. cont. \\ $\epsilon = 0.8$ \\ $\sigma_1=0.5$ \\$\sigma_2 = 20$} & -0.39 \\ \hline
  4 targets 2 clutters & 1 target masked & 3 dB& -4.77 \\ \hline
  4 targets 2 clutters & detected & 6 dB& -6.02 \\ \hline
  4 targets 2 clutters & no detection & \parbox[t]{1.4cm}{\fontsize{5}{4}\selectfont eps. cont. \\ $\epsilon = 0.9$ \\ $\sigma_1=0.25$ \\$\sigma_2 = 10$} & -0.23 \\ \hline
  4 targets 2 clutters & no detection & \parbox[t]{1.4cm}{\fontsize{5}{4}\selectfont eps. cont. \\ $\epsilon = 0.8$ \\ $\sigma_1=0.5$ \\$\sigma_2 = 20$} & -0.12 \\ \hline
  1 target 3 clutters &	detected & 3 dB& -5.54 \\ \hline
  1 target 3 clutters &	detected & 6 dB& -5.73 \\ \hline
  1 target 3 clutters &	no detection & \parbox[t]{1.4cm}{\fontsize{5}{4}\selectfont eps. cont. \\ $\epsilon = 0.9$ \\ $\sigma_1=0.25$ \\$\sigma_2 = 10$} & -0.85 \\ \hline
  1 target 3 clutters &	no detection & \parbox[t]{1.4cm}{\fontsize{5}{4}\selectfont eps. cont. \\ $\epsilon = 0.8$ \\ $\sigma_1=0.5$ \\$\sigma_2 = 20$} & -0.33 \\ \hline
  \hline
\end{tabular} }
\end{center}
\end{table}

\begin{table}[ht!]
\begin{center}
\caption{Environment setup for Fig. 3, 4, 5 and 6. Contaminated Gaussian noise with $\epsilon=0.9$, $\sigma_1 = 0.25$ and $\sigma_2 = 10$ is used.}
\scalebox{0.8} {
\label{tab:3}
\begin{tabular}{|l|l|l|l|l|}
\hline
  & x-axis (km) & y-axis (km) & \parbox[t]{1.34cm}{Doppler \\ Shift (Hz)} & K \\ \hline
  transmitter & 0 & 10 & - & - \\ \hline
  receiver & 0 & 0 & - & - \\ \hline
  $target_1$ & 10 & 0 & 200 & 1 \\ \hline
  $target_2$ & 20 & 0 & 157 & 1 \\ \hline
  $clutter_1$ & 28 & 33 & 0 & 1 \\ \hline
  \hline
\end{tabular} }
\end{center}
\end{table}

\begin{figure}[h!]
\label{fig:3}

  \centering
    \includegraphics[width=8cm, height=35mm]{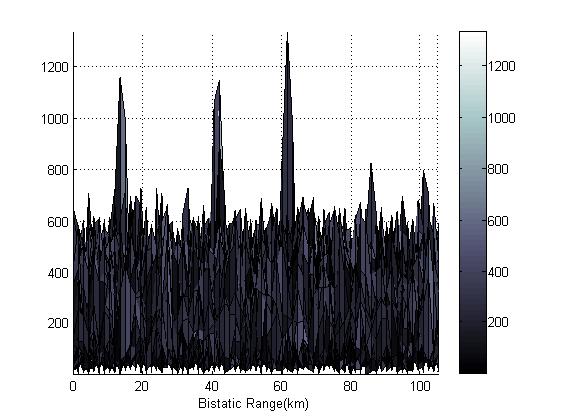}
    \caption{Bistatic range-cut plot for contaminated Gaussian noise using NFFT in Equation (12a). Bistatic range information for all obstacles are easily detectable.}
\end{figure}
\begin{figure}[h!]
\label{fig:3}

  \centering
    \includegraphics[width=8cm, height=35mm]{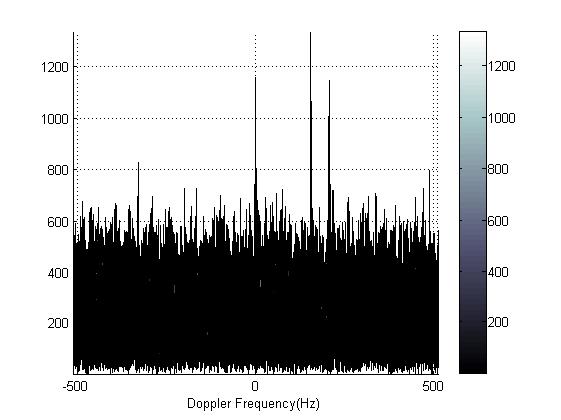}
    \caption{Doppler-cut plot for contaminated Gaussian noise using NFFT in Equation (12a). Bistatic range information for all obstacles are easily detectable.}
\end{figure}
\begin{figure}[h!]
\label{fig:3}

  \centering
    \includegraphics[width=8cm, height=35mm]{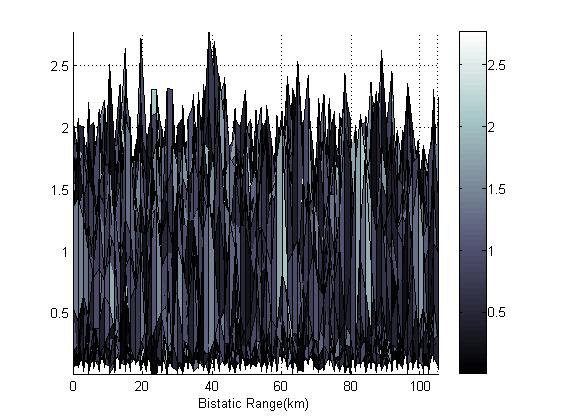}
    \caption{Bistatic range-cut plot for contaminated Gaussian noise using FFT in Equation (11). All targets and clutters are masked.}
\end{figure}
\begin{figure}[h!]
\label{fig:3}

  \centering
    \includegraphics[width=8cm, height=35mm]{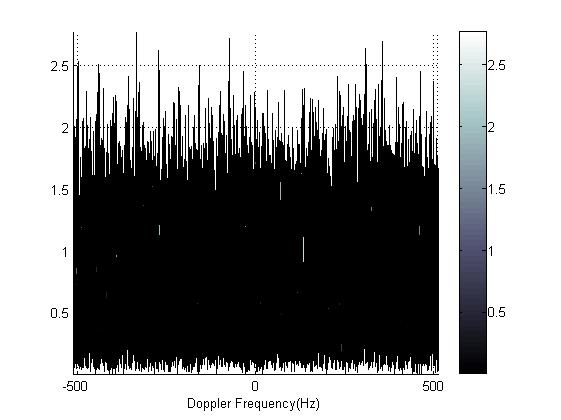}
    \caption{Bistatic range-cut plot for contaminated Gaussian noise using FFT in Equation (11). All targets and clutters are masked.}
\end{figure}

\vfill\newpage
\vfill\newpage

%\section{REFERENCES}
%\label{sec:refs}
%
%List and number all bibliographical references at the end of the
%paper. The references can be numbered in alphabetic order or in
%order of appearance in the document. When referring to them in the
%text, type the corresponding reference number in square brackets
%as shown at the end of this sentence \cite{C2}. \textbf{An
%additional final page (the fifth page, in most cases) is allowed,
%but must contain only references to the prior literature.}

\bibliographystyle{IEEEbib}
\bibliography{strings}
\end{document}